\documentclass[fleqn]{article}
\usepackage{amsmath,amssymb}
\usepackage{epsfig,graphicx}

\begin{document}
\title{\vskip-1.7cm \bf CFT driven cosmology and the DGP/CFT
correspondence}
\date{}
\author{A.O.Barvinsky$^{1,\,2}$, C.Deffayet$^3$ and
A.Yu.Kamenshchik$^{4,\,5}$}
\maketitle \hspace{-8mm}
{\,\,$^{1}$\em
Theory Department, Lebedev
Physics Institute, Leninsky Prospect 53, Moscow 119991, Russia\\
$^{2}$LMPT CNRS-UMR 6083, Univ. of Tours, Parc de Grandmont,
37200 Tours, France\\
$^{3}$APC UMR 7164 (CNRS, Univ. Paris 7, CEA, Obs. de Paris), 12 rue
Alice Domon et L\'eonie Duquet, 75205 Paris Cedex 13, France\\
$^{4}$Dipartimento di Fisica and INFN, via Irnerio 46, 40126
Bologna, Italy\\
$^{5}$L.D.Landau Institute for Theoretical Physcis of Russian
Academy of Sciences, Kosygin str. 2, 119334 Moscow, Russia\\}

\begin{abstract}
We present a dual 5D braneworld picture of a recently suggested
model for a microcanonical description of a 4D cosmology driven by a
conformal field theory with a large number of quantum fields. The 5D
side of the duality relation is represented by a generalized brane
induced gravity  model in a Schwarzschild-de Sitter bulk. The values
of the bulk cosmological and the induced 4D cosmological constants
are determined by requiring the absence of conical singularity at
the de Sitter horizon of the Euclidean Schwarzschild-de Sitter bulk.
Those values belong to the vicinity of the upper bound of a range of
admissible values for the cosmological constant. This upper bound is
enforced by the 4D CFT and coincides with the natural gravitational
cutoff in a theory with many quantum species. The resulting DGP/CFT
duality suggests the possibility of a new type of {\em background
independent} correspondence. A mechanism for inverting the sign of
the effective cosmological constant is found, which might reconcile
a negative value of the primordial cosmological constant compatible
with supersymmetry with the one required by inflationary cosmology.

\end{abstract}
\maketitle
\section{Introduction}
As is widely recognized, the AdS/CFT correspondence \cite{AdS/CFT}
is an efficient tool for the nonperturbative analysis of various
problems, providing a new insight in the underlying physical
mechanisms and relating classical and quantum domains in relevant
weak and strong coupling regimes. It is natural, therefore, to look
for new applications where the hypothesis of this (or a similar)
correspondence can work and bring new nonperturbative results. This
becomes even more interesting in the context of a recently suggested
relation between the cutoff in classical black hole physics and the
properties of a particle model \cite{species1,species2},
establishing the dependence of this cutoff on the number of quantum
species $N$ in the theory. The AdS/CFT correspondence relates dual
models in the domains of couplings separated by this cutoff and,
thus, directly involves this concept.

Together with the hierarchy problem of the electroweak vs Planckian
scales \cite{species1}, one of the most intriguing implications of
these ideas concerns problems in early and late cosmology ranging
from the origin of inflationary Universe to the dark energy. In this
sense, a recently suggested cosmological model driven by a conformal
field theory (CFT) with a large number of conformal fields
\cite{slih,why} is a perfect tool for such a duality analysis. This
model is of a particular interest, because it incorporates a
microcanonical initial conditions in cosmology, capable of resolving
the string landscape problem, and possibly contains a mechanism for
the cosmological acceleration \cite{why,boost}.

In this paper we show that this essentially quantum four-dimensional
(4D) model has a dual description in terms of a classical 5D theory
incorporating a dynamically evolving 4D brane with brane induced
gravity in a Schwarzschild-{\em de Sitter} 5D bulk (what we will
call here the generalized DGP model). The parameters of this model
--- its cosmological constant $\Lambda_5$ and the black hole mass
belong to the classical domain much below the cutoff, whereas the 4D
cosmology of \cite{slih} on the CFT side of the duality relation
lies close to the upper boundary of the admissible range of values
for the 4D cosmological constant $\Lambda_4$. This range was derived
in \cite{slih} and suggested to be of interest for dealing with the
string landscape and the cosmological constant problem. Distinctive
peculiarities of this {\em DGP/CFT correspondence} is that it is
realized not on the AdS background, but on the Schwarzschild-{\em de
Sitter} one, and does not resort to group-theoretical and, in
particular, supersymmetry arguments. Moreover, the parameters of the
system (that is its cosmological constant $\Lambda_5=2\Lambda_4$,
temperature and the radiation energy or the black hole mass on the
5D side) are not freely specifiable, but are entirely determined in
terms of the Planck mass by the particle content of the model.

The organization of the paper is as follows. In Sect.2 we
recapitulate the construction of the CFT driven cosmology of
\cite{slih,why}, its initial conditions prescribed by the
microcanonical density matrix and late time evolution incorporating
the cosmological acceleration in the form of the Big Boost scenario
\cite{boost}. In Sect.3 we discuss the single branch generalized DGP
model with a positive $\Lambda_5$ in the bulk, which for the vacuum
case was observed in \cite{boost} to be equivalent to the CFT
cosmology. Sect.4 presents the derivation of this equivalence in the
Euclidean signature spacetime which underlies the 4D cosmological
instantons dominating the statistical sum path integral. On the 5D
side these instantons represent a dynamically evolving brane
embedded into the Schwarzschild-de Sitter bulk with
$\Lambda_5=2\Lambda_4$, its black hole mass being related to the
thermal radiation on the brane. Complete duality of 4D and 5D
descriptions implies the specification of the cosmological constant
and the radiation energy, which on the 4D side follows from the
quantum Einstein equations and on the 5D side is enforced by the
requirement of the absence of conical singularity in the bulk
geometry. Together these conditions  select from the finite range of
$\Lambda_4$, admissible from the 4D viewpoint, a value close
to the upper limit of this range, where the gravitating CFT belongs
to a strong coupling regime treated within $1/N$-expansion. This
property is derived in Sect.5 for a special set of static Einstein
instantons of the CFT cosmology and for the non-static solutions
called {\em garlands} in \cite{slih}
--- the instantons giving rise to expanding inflationary
models. Sect.6 contains an attempt to generalize this DGP/CFT
correspondence to the case of the {\em background independent}
formulation, when the field configuration is not limited by symmetry
or weak field linearization restrictions --- an idea suggested in
\cite{DvaliGomez} on the basis of quantum information and gravity
cutoff bounds. In particular, it is shown that a full nonlinear
trace part of effective Einstein equations can be exactly the same
on both sides of the duality relation. This turns out to be true
under the assumption of a particle phenomenology motivated by the
superconformal ${\cal N}=4$ Yang-Mills theory which underlies the
$AdS_5\times S_5/SCFT$ correspondence of \cite{AdS/CFT}. In Sect.7
we discuss an interesting mechanism based on the conformal anomaly
to reverse the sign of the primordial negative cosmological
constant. This could reconcile supersymmetry with inflation. Sect.8
contains conclusions and a brief discussion of the results.

\section{CFT driven cosmology}

The initial conditions for the CFT driven cosmology of
\cite{slih,why} were prescribed in terms of {\bf a} microcanonical
density matrix. They are based on the corresponding statistical sum
for a spatially closed cosmology ($S^3$-topology of spatial
sections). It is represented by the path integral over the periodic
scale factor $a(\tau)$ and lapse function $N(\tau)$ of the
minisuperspace metric
    \begin{eqnarray}
    ds^2 = N^2(\tau)\,d\tau^2+a^2(\tau)\,d^2\Omega^{(3)} \label{FRW}
    \end{eqnarray}
on the toroidal spacetime of $S^1\times S^3$ topology
    \begin{eqnarray}
    e^{-\varGamma}=\!\!\int\limits_{\,\,\rm periodic}
    \!\!\!\! D[\,a,N\,]\;
    e^{-\varGamma_E[\,a,\,N\,]}.   \label{1}
    \end{eqnarray}
Here $\varGamma_E[\,a,\,N]$ is the Euclidean effective action of all
inhomogeneous ``matter" fields on the minisuperspace background
(\ref{FRW}) including also the Einstein action with with the
renormalized gravitational $G$ and cosmological $\Lambda$ constants.
This action incorporates the integration over all degrees of freedom
of the system not contained in its minisuperspace sector.

Under the assumption that the system is dominated by free matter
fields conformally coupled to gravity this action is exactly
calculable by the conformal transformation converting (\ref{FRW})
into the static Einstein metric with $a={\rm const}$ \cite{slih}. In
units of the Planck mass $m_P=(3\pi/4G)^{1/2}$ the action reads
    \begin{eqnarray}
    &&\varGamma_E[\,a,N\,]=m_P^2\int d\tau\,N \left\{-aa'^2
    -a+ \frac\Lambda3 a^3\right.\nonumber\\
    &&\qquad\qquad\qquad\qquad\qquad\qquad
    \left.+\,B\left(\frac{a'^2}{a}
    -\frac{a'^4}{6 a}\right)+\frac{B}{2a}\,\right\}
    +F(\eta),                              \label{effaction}\\
    &&F(\eta)=\pm\sum_{\omega}\ln\big(1\mp
    e^{-\omega\eta}\big),\,\,\,\,\,
    \eta=\int \frac{d\tau N}a,                    \label{freeenergy}
    \end{eqnarray}
where $a'\equiv da/Nd\tau$. The first three terms in curly brackets
of (\ref{effaction}) represent the Einstein action with a primordial
(but renormalized by quantum corrections) cosmological constant
    \begin{equation}
    \Lambda\equiv 3H^2
    \end{equation}
($H$ is the corresponding Hubble constant), the $B$-terms correspond
to the contribution of the conformal anomaly \cite{anomalyaction}
and the contribution of the vacuum (Casimir) energy $(B/2a)$ of
conformal fields on a static Einstein spacetime. $F(\eta)$ is the
free energy of these fields -- a typical boson or fermion sum over
field oscillators with energies $\omega$ on a unit 3-sphere, $\eta$
playing the role of the inverse temperature --- an overall
circumference of the toroidal instanton measured in units of the
conformal time. The constant $B$,
    \begin{eqnarray}
    B=\frac{3\beta}{4 m_P^2}=\frac{\beta G}\pi,   \label{B}
    \end{eqnarray}
is determined by the coefficient $\beta$ of the topological
Gauss-Bonnet invariant $E = R_{\mu\nu\alpha\gamma}^2-4R_{\mu\nu}^2 +
R^2$ in the overall conformal anomaly of quantum fields
    \begin{equation}
    g_{\mu\nu}\frac{\delta
    \varGamma_A}{\delta g_{\mu\nu}} =
    \frac{1}{4(4\pi)^2}g^{1/2}
    \left(\alpha \Box R +
    \beta E + \gamma C_{\mu\nu\alpha\beta}^2\right)    \label{anomaly}
    \end{equation}
Here $\varGamma_A$ is the CFT action in the external gravitational
field generating the conformal anomaly and $C^2_{\mu\nu\alpha\beta}$
is the Weyl tensor squared term. For a model with $N_0$ scalars,
$N_{1/2}$ Weyl spinors and $N_{1}$ gauge vector fields it reads
\cite{Duffanomaly}
    \begin{eqnarray}
    \beta=\frac1{360}\,\big(2 N_0+11 N_{1/2}+
    124 N_{1}\big).                \label{100}
    \end{eqnarray}

The status of other coefficients of (\ref{anomaly}) is as follows.
The coefficient $\gamma$ does not contribute to (\ref{effaction})
because the Weyl tensor vanishes for any FRW metric. A nonvanishing
$\alpha$ induces higher derivative terms $\sim \alpha (a'')^2$ in
the action and, therefore, adds one extra degree of freedom to the
minisuperspace sector of $a$ and $N$ which generically leads to
instabilities\footnote{In Einstein theory this sector does not
contain physical degrees of freedom at all, which solves the problem
of the formal ghost nature of $a$ in the Einstein Lagrangian.
Addition of higher derivative term for $a$ does not necessarily lead
to a ghost -- the additional degree of freedom can have a good sign
of the kinetic term as it happens in $f(R)$-theories, but still
leads to the instabilities discovered in \cite{Starobinsky}.}. But
$\alpha$ can be renormalized to zero by adding a finite {\em local}
counterterm $\sim R^2$ admissible by the renormalization theory.
Such a {\em number of degrees of freedom preserving} renormalization
was assumed in \cite{slih} to keep the theory consistent both at the
classical and quantum levels. It is important that this finite
renormalization changes the value of the Casimir energy of conformal
fields in closed Einstein cosmology in such a way that for all spins
this energy is universally expressed in terms of the same conformal
anomaly coefficient $B$ (corresponding to the $B/2a$ term in
(\ref{effaction})) \cite{universality,slih}\footnote{This
universality property follows from the fact that in a conformally
flat spacetime --- the case of a decompactified universe $R\times
S^3$ --- the Casimir energy is entirely determined by the conformal
anomaly coefficients \cite{universality}. We thank P.Mazur and
E.Mottola for the discussion of this point.}.

The path integral (\ref{1}) is dominated by saddle points ---
solutions of the equation $\delta\varGamma_E/\delta N(\tau)=0$ which
reads as
    \begin{eqnarray}
    &&-\frac{a'^2}{a^2}+\frac{1}{a^2}
    -\frac{B}2 \left(\,-\frac{a'^2}{a^2}
    +\frac1{a^2}\right)^2 =
    \frac\Lambda3+\frac{\cal C}{ a^4},     \label{efeq}
    \end{eqnarray}
with $\cal C$ given by
    \begin{eqnarray}
    &&{\cal C} = \frac1{m_P^2}\frac{dF(\eta)}{d\eta}=
    \frac1{m_P^2}
    \sum_\omega\frac{\omega}{e^{\omega\eta}\mp 1},\,\,\,\,
    \eta = \oint\frac{d\tau}{a}\equiv 2k \int_{\tau_-}^{\tau_+}
    \frac{d\tau}{a}.                       \label{bootstrap}
    \end{eqnarray}
This Euclidean Friedmann equation is modified by the anomalous
$B$-term and the radiation term ${\cal C}/a^4$. The constant ${\cal
C}$ sets the amount of radiation and determines the energy of the
gas of thermally excited particles with the inverse temperature
$\eta$ -- the instanton period in units of the conformal time. The
latter is given in (\ref{bootstrap}) by the integral over the full
period of $\tau$ or the $2k$-fold integral between the two
neighboring turning points of the scale factor history $a(\tau)$,
$\dot a(\tau_\pm)=0$. This $k$-fold nature implies that in the
periodic solution the scale factor oscillates $k$ times between its
maximum and minimum values $a_\pm=a(\tau_\pm)$, $a_-\leq a(\tau)\leq
a_+$,
    \begin{eqnarray}
    a^2_\pm=
    \frac1{2H^2}\Big(1\pm\sqrt{1-2BH^2-4H^2{\cal C}}\Big). \label{apm}
    \end{eqnarray}
Thus, the period of the solutions is determined as a function of $G$
and $\Lambda$ from the second of Eqs.(\ref{bootstrap}) and is not
freely specifiable. This is the artifact of a microcanonical
ensemble (see \cite{why}) with only two freely specifiable
dimensional parameters
--- the renormalized gravitational and renormalized cosmological
constants.

\begin{figure}[h]
\centerline{\epsfxsize 8.5cm \epsfbox{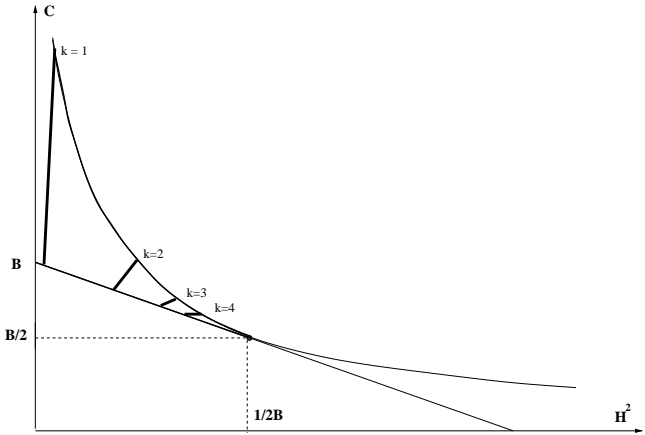}} \caption{\small The
instanton domain in the $(H^2,C)$-plane with four garland families
$k=1,2,3,4$. $C={\cal C}+B/2$ is the amount of radiation constant
including the contribution of the Casimir energy.
 \label{Fig.1}}
\end{figure}

As shown in \cite{slih}, such solutions represent $S^3\times S^1$
garland-type instantons which exist only in the limited range of the
cosmological constant $\Lambda=3H^2$
    \begin{eqnarray}
    0<H^2_{\rm min}<H^2<
    \frac{\pi}{2\beta G}=\frac1{2B}.                \label{landscape}
    \end{eqnarray}
In this range they form an infinite $k=0,1,2,...$ sequence of
one-parameter families depicted in Fig.\ref{Fig.1} in the
two-dimensional plane of the Hubble constant $H^2$ and the amount of
radiation constant $C={\cal C}+B/2$ (including together with the
energy of the radiation gas $\cal C$, see (\ref{bootstrap}), also
the Casimir energy constant $B/2$.) These families interpolate
between the two boundaries of a curvilinear triangle of the
instanton domain in the $(H^2,C)$-plane -- the lower straight line
boundary $C=B-B^2H^2$ and the upper hyperbolic boundary $C=1/4H^2$.
Their sequence at $k\to\infty$ accumulates at the corner of this
triangle --- the upper bound of the range (\ref{landscape})
$H^2=1/2B$.

This bound represents a new quantum gravity scale which tends to
infinity when one switches the quantum effects off, $\beta\to 0$.
The lower bound $H^2_{\rm min}$
--- the lowest point of $k=1$ family --- can be obtained numerically
for any field content of the model. For a large number of conformal
fields $N$, and therefore a large $\beta\sim N$, the lower bound is
of the order $H^2_{\rm min}\sim 1/\beta G$. Thus the restriction
(\ref{landscape}) suggests a kind of $1/N$ solution of the
cosmological constant problem, because specifying a sufficiently
high number of conformal fields one can achieve a primordial value
of $\Lambda$ well below the Planck scale where the effective theory
applies, but high enough to generate a sufficiently long
inflationary stage. Also this restriction can be potentially
considered as a selection criterion for the landscape of string
vacua \cite{slih,why}.

The solutions of the system (\ref{efeq})-(\ref{bootstrap}) include
the vacuum Hartle-Hawking instantons of \cite{HH} (see their dual
braneworld description in \cite{HHR}) with no radiation ${\cal
C}=0$. They represent the Euclidean de Sitter spacetime with the
effective Hubble factor
    \begin{eqnarray}
    H^2_{\rm eff}=\frac{1-\sqrt{1-2BH^2}}B,   \label{HHHeff}
    \end{eqnarray}
corresponding to the degenerate case when a torus $S^3\times S^1$
gets ripped at the vanishing value of the scale factor $a_-=0$ and
topologically becomes a 4-sphere $S^4$ --- a purely vacuum
contribution to the statistical sum. The vacuum nature of these
instantons follows from the fact that their conformal time period in
(\ref{bootstrap}) is divergent at $\tau_-$ in view of $a_-=0$ and
generates zero effective temperature $\sim1/\eta$ with $F(\eta)=0$.
Such solutions exist for all $H^2\leq 1/2B$ (the horizontal segment
at $C\equiv{\cal C}+B/2=B/2$ in Fig.\ref{Fig.1}), but they are ruled
out in the statistical sum by their infinite {\em positive}
effective action. This property is due to the contribution of the
conformal anomaly (cf. $1/a$-dependence in the kinetic $B$-terms of
the effective action (\ref{effaction}))\footnote{Note that on the
vacuum solution of (\ref{efeq}) $a'^2(\tau_-)=1$, and the integrand
of (\ref{effaction}) tends to $+\infty$ at $\tau_-$ with $a\to 0$.
In its turn, the absence of a conical singularity at the cap of the
$S^4$ instanton follows from a special UV renormalization which
eradicates higher-order derivatives from the effective Friedmann
equation and simultaneously renders a particular value of the
Casimir energy $B/2$ \cite{slih}.}. Hence the tree-level predictions
of the theory with a negative Euclidean action are drastically
changed by the effect of the conformal anomaly.

The solutions of (\ref{efeq})-(\ref{bootstrap}) also include the
exactly static instantons with $a_-=a_+=1/H\sqrt2$ which occupy the
full upper hyperbolic boundary of the instanton domain in
Fig.\ref{Fig.1}. They exist for all $H^2$ from the range
(\ref{landscape}), but they are not interesting from the viewpoint
of the early Universe, because they stay static and do not give rise
to expanding cosmological models. As they serve as a simple working
tool for the DGP/CFT selection rules, they will be considered in
much detail in Sect.5 below.

The quadratic equation (\ref{efeq}) can be solved for the Euclidean
Hubble factor squared $(1-a'^2)/a^2$. For periodic instantons of the
above type only one root is admissible\footnote{Only this sign of
the square root in (\ref{FriedmannE}) is possible because otherwise
the Euclidean solutions with $a'^2\geq 0$  exist only outside of the
domain $a_-<a<a_+$ and violate the periodicity requirement imposed
by the construction of the path integral for the statistical sum.},
which gives
    \begin{eqnarray}
    &&-\frac{a'^2}{a^2}+\frac{1}{a^2}=
    \frac1B\left\{1-
    \sqrt{1-2B\left(\frac\Lambda3
    +\frac{\cal C}{a^4}\right)}\right\}\,.         \label{FriedmannE}
    \end{eqnarray}

This equation obviously recovers the Einstein theory in the limit of
a small $B$ and low matter density $\varepsilon\equiv (3/8\pi
G)(\Lambda/3+{\cal C}/a^4)$, $GB\varepsilon\ll 1$, because its right
hand side takes in this limit a usual GR form $8\pi G\varepsilon/3$
independent of $B$. Also it shows that the Casimir energy that was
initially present in the effective action (\ref{effaction}), the
term $a^3\times B/2a^4$, does not weigh. In a static Einstein
Universe this energy component has the same equation of state as
radiation \cite{E_0}, but in contrast to the real radiation with the
Planck (or Bose-Fermi) spectrum (\ref{bootstrap}) it does not enter
the right hand side of (\ref{FriedmannE}). This degravitation is a
result of the number of degrees of freedom preserving
renormalization mentioned above.

The gravitational instantons of the above type serve as a source of
initial conditions for the cosmological evolution in the physical
spacetime. Lorentzian signature Universes nucleate from the minimal
surface of these instantons at $\tau_+$ --- the point of the maximal
expansion of their Euclidean solutions. The latter when analytically
continued to the complex plane by the rule $\tau=\tau_++it$ give the
evolution in real Lorentzian time $t$. With $\dot a\equiv da/dt=ia'$
the generalized Friedmann equation (\ref{FriedmannE}) for this
evolution takes the form in the Lorentzian regime
    \begin{eqnarray}
    &&\frac{\dot a^2}{a^2}+\frac{1}{a^2}=
    \frac1B\left\{1-
    \sqrt{1-2B\left(\frac\Lambda3
    +\frac{\cal C}{a^4}\right)}\right\}.         \label{FriedmannL}
    \end{eqnarray}

As shown in \cite{boost}, this equation first admits the stage of
inflationary (quasi-exponential) expansion driven by $\Lambda$ and
terminating with the the usual exit scenario when $\Lambda$ is a
decaying composite field (inflaton). During this stage a particle
production of a conformally non-invariant matter takes over the
polarization effects of conformal fields. After the thermalization
this matter gives rise to an energy density $\varepsilon$ which
replaces in (\ref{FriedmannL}) the energy density of the primordial
cosmological constant and radiation, $\Lambda/3+{\cal C}/a^4\to 8\pi
G\varepsilon/3$. Then (\ref{FriedmannL}) takes the form
    \begin{eqnarray}
    \frac{\dot a^2}{a^2}+\frac{1}{a^2}=
    \frac\pi{\beta G}\left\{\,1-
    \sqrt{\,1-\frac{16 G^2}3\,
    \beta\varepsilon}\,\right\},              \label{modFriedmann}
    \end{eqnarray}
where we expressed $B$ according to (\ref{B}) and, due to square
root non-analyticity, incorporates the so-called {\em big boost}
scenario \cite{boost}. In the vicinity of a vanishing square root
argument the deceleration of the Universe can go over into
acceleration followed by a big boost singularity at $1-16\beta
G^2\varepsilon/3=0$: at a finite value of the Hubble factor
cosmological acceleration becomes infinite\footnote{This singularity
should be regulated by a certain UV completion of the model or
suppressed by the $R^2$-term in the action, disregarded above in the
$1/N$-approximation. Eradication of $R^2$-terms by the number of
degrees of freedom preserving renormalization applies in the large
CFT sector of the theory, but this does not prohibit such terms in
other sectors of the full model. We thank A.A.Starobinsky for this
observation.}, $\ddot a\to +\infty$. This branching point of Eq.
(\ref{modFriedmann}) can be reached for a decreasing $\varepsilon$
if $\beta$ is promoted to the level of an adiabatically growing
variable -- the mechanism of evolving number of light Kaluza-Klein
or winding modes, induced by evolving non-stabilized extra
dimensions \cite{why,boost}.

\section{The generalized single-branch DGP cosmology}
The fact that the CFT driven cosmology of the above type has a dual
braneworld description was first observed in \cite{boost}. The dual
picture is given by a generalized DGP model \cite{DGP} including
together with 4D and 5D Einstein-Hilbert terms also a 5D
cosmological constant, $\Lambda_5$, in the special case of a {\em
vacuum} state with a vanishing matter density on the brane.
Interestingly, this vacuum DGP cosmology {\em exactly} corresponds
to the model of \cite{slih} with the 4D cosmological constant
$\Lambda$ simulated by the 5D cosmological constant $\Lambda_5$.

Indeed, in this model (provided one neglects the bulk curvature),
gravity interpolates between a 4D behaviour at small distances and a
5D behaviour at large distances, with the crossover scale between
the two regimes being given by $r_c$,
    \begin{eqnarray}
    \frac{G_5}{2G}=r_c,      \label{DGPscale}
    \end{eqnarray}
and in the absence of stress-energy exchange between the brane and
the bulk, the modified Friedmann equation takes the form
\cite{DGPDeffayet}
    \begin{eqnarray}
    \frac{\dot a^2}{a^2}+\frac{1}{a^2}-
    r_c^2 \left(\,\frac{\dot a^2}{a^2}
    +\frac{1}{a^2}-\frac{8\pi G}3\,\rho\right)^2 =
    \frac{\Lambda_5}{6}
    +\frac{{\cal C}}{ a^4}.             \label{FriedmannDGP}
    \end{eqnarray}
Here ${\cal C}$ is the integration of the bulk Einstein's equation,
which corresponds to a nonvanishing Weyl tensor in the bulk (or a
mass of the 5D Schwarzschild black hole)
\cite{BinDefLan,bulkBH,Kraus} and $\rho$ is the energy density on
the brane. This equation with $\rho=0$ exactly coincides with the
Lorentzian version of the effective Friedmann equation in the CFT
cosmology (\ref{efeq}) under the identifications
    \begin{eqnarray}
    &&B\equiv\frac{\beta G}\pi=2 r_c^2, \label{1000}\\
    &&\Lambda=\frac{\Lambda_5}2.        \label{1001}
    \end{eqnarray}
These identifications imply that in the DGP limit $G\ll r_c^2$, the
anomaly coefficient $\beta$ is much larger than 1, and this looks
very much like the generation of the vacuum DGP model for any value
of the dark radiation ${\cal C}/a^4$ from the CFT cosmology with a
very large $\beta\sim m_P^2 r_c^2\gg 1$.

However, there are several differences. First, the CFT driven DGP
model does not incorporate the self-accelerating branch
\cite{DGPDeffayet,DDG} of the conventional DGP cosmology. This
corresponds to the fact that only one sign of the square root is
admissible in Eq.(\ref{FriedmannL}) --- a property dictated by the
instanton initial conditions at the nucleation of the Lorentzian
spacetime from the Euclidean one. So, one does not have to worry
about possible instabilities associated with the self-accelerating
branch \cite{instabilityofacceleratingbranch}.

Another difference concerns the way the matter energy density
manifests itself in the Friedmann equation for the non-vacuum case.
In the CFT model it enters the right hand side of the equation as a
result of the decay of the effective 4D cosmological constant
$\Lambda$ resulting in (\ref{modFriedmann}). In the conventional DGP
model it appears inside the parenthesis of the left hand side of
equation (\ref{FriedmannDGP}), so that the DGP Hubble factor reads
as
    \begin{eqnarray}
    \frac{\dot a^2}{a^2}+\frac{1}{a^2}=
    \frac{8\pi G}3\,
    \rho+
    \frac1{2r_c^2}\left\{\,1-
    \sqrt{\,1-4r_c^2
    \left(\frac{\textstyle\Lambda_5}{\textstyle 6}
    +\frac{\textstyle\cal C}{\textstyle a^4}-
    \frac{\textstyle 8\pi G}{\textstyle 3}\,
    \rho\right)}\,\right\}                     \label{modFriedmannDGP}
    \end{eqnarray}
(note the negative sign of $\rho$ under the square root and the
extra first term on the right hand side).

Due to the presence of the cosmological constant in the bulk this
generalized DGP model also admits the big boost acceleration
scenario \cite{boost}. Indeed, for positive $\Lambda_5$ satisfying a
very weak bound
    \begin{eqnarray}
    \Lambda_5>\frac3{2r_c^2} \label{bound}
    \end{eqnarray}
Eq.(\ref{modFriedmannDGP}) has a solution for which,  during the
cosmological expansion with $\rho\to 0$, ${\cal C}/a^4\to 0$, the
argument of the square root vanishes and the acceleration tends to
infinity. For that to happen, one does not actually need a growing
$r_c$ or $\beta$, like in the CFT driven model above (see the end of
the previous section). The DGP crossover scale $r_c$ can be
constant, and the big boost singularity will still occur provided
the lower bound (\ref{bound}) is satisfied. For a positive
$\Lambda_5$ below this bound, the acceleration stage is eternal with
an asymptotic value of the Hubble factor given by
    \begin{eqnarray}
    \left.\frac{\dot a^2}{a^2}\,\right|_{\;a\to\infty}\to\,
    \frac{1-\sqrt{1-2r_c^2\Lambda_5/3}}{2r_c^2},\,\,\,\,\,\,
    0<\Lambda_5\leq\frac3{2r_c^2},
    \end{eqnarray}
whereas for $\Lambda_5\leq 0$ the model bounces from some maximal
value of a scale factor and starts recollapsing.

\section{The generalized Euclidean DGP model}
The Euclidean action of the {\bf generalized} {\em vacuum} DGP model of the
above type reads
    \begin{eqnarray}
    &&S_{DGP}[\,G_{AB}(X)]=
    -\frac1{16\pi G_5}\int\limits_{\bf B} d^5X\,
    G^{1/2}\,\Big(R^{(5)}(\,G_{AB}\,)-2\Lambda_5\Big)
    \nonumber\\
    &&\qquad\qquad\qquad\qquad
    -\int\limits_{\bf b} d^4x\,g^{1/2}\left(\frac1{8\pi G_5}[\,K\,]
    +\frac1{16\pi G_4}\,R(\,g_{\mu\nu}\,)
    \right).     \label{10.25}
    \end{eqnarray}
Here the 5D bulk ${\bf B}$ is $Z_2$ symmetric with respect to the
brane $\bf b$ bearing the 4D Einstein action for the induced metric
$g_{\mu\nu}(x)$ and the Gibbons-Hawking term for the bulk Einstein
action for the metric $G_{AB}(X)$. The latter involves the jump of
the trace of the extrinsic curvature across the brane $[K]=2K$.
Variational equations for this action include the Einstein equations
in the bulk and the Israel junction conditions on the brane
    \begin{eqnarray}
    &&R^{(5)}_{AB}-\frac12\,G_{AB}\,R^{(5)}=
    -\Lambda_5\,G_{AB},                          \label{bulkEinstein}\\
    &&\big(K_{\mu\nu}-K g_{\mu\nu}\big)_{\bf b}=
    4\pi G_5\,S_{\mu\nu},                        \label{Israel}
    \end{eqnarray}
where $S_{\mu\nu}$ is a surface stress tensor on the brane
coinciding with the 4D Einstein tensor
    \begin{eqnarray}
    S_{\mu\nu}=-\frac1{8\pi G_4}\left(
    R_{\mu\nu}
    -\frac12 \,g_{\mu\nu}\,R \right).        \label{surfacestress}
    \end{eqnarray}

If we apply these equations to the spherically symmetric ansatz for
the 5D metric, then we get the Schwarzschild-de Sitter solution with
the Schwarzschild $R_S$ and de Sitter horizon $R_{dS}$ radii
    \begin{eqnarray}
    &&ds_{(5)}^2=f(R)dT^2+
    \frac{dR^2}{f(R)}+R^2d\Omega_{(3)}^2,  \label{SdS}\\
    &&f(R)=1-H^2R^2-\frac{R_S^2}{R^2},\\
    &&H^2=\frac1{R_{dS}^2}=\frac{\Lambda_5}6,
    \end{eqnarray}
with the spherical brane embedded into the bulk according to the
equations
    \begin{eqnarray}
    &&R=a(\tau),\\
    &&T=T(\tau).
    \end{eqnarray}
Here  $\tau$ is the proper time on the brane which has the induced
FRW metric (\ref{FRW}) with the lapse $N=1$ and the scale factor
$a(\tau)$, $ds_{(4)}^2=d\tau^2+a^2(\tau)d\Omega_{(3)}^2$. This
embedding implies the evolution law for the 5D time coordinate
$T(\tau)$ ($T'(\tau)\equiv dT(\tau)/d\tau$)
    \begin{eqnarray}
    T'(\tau)=
    \frac{\sqrt{f(a)-a'^2}}{f(a)}   \label{dTdtau}
    \end{eqnarray}
and the following expression for the outward pointing normal to the
boundary of the domain $R\geq a(\tau)$ \cite{Kraus}
    \begin{eqnarray}
    n^A=\left(\frac{a'}f,\,-\sqrt{f(a)-a'^2},\,0,\,0,\,0\right).
    \end{eqnarray}
The nonvanishing components of the respective extrinsic curvature
$K_{\mu\nu}=\nabla_\mu n_\nu$ have the form
    \begin{eqnarray}
    &&K_{ij}=-\frac{\sqrt{f-a'^2}}a\,g_{ij},\\
    &&K_{\tau\tau}=-\frac1{a'}\frac{d}{d\tau}\sqrt{f-a'^2},
    \end{eqnarray}
where $g_{ij}=a^2(\tau)\gamma_{ij}$ is a metric of the 3-sphere of
the radius $a(\tau)$.

Since
    \begin{eqnarray}
    &&R^{(4)}_{\tau\tau}-\frac12 g_{\tau\tau}R^{(4)}=
    3\left(\frac{a'^2}{a^2}-\frac1{a^2}\right),\\
    &&R_{ij}^{(4)}-\frac12 g_{ij}R^{(4)}=-g_{ij}
    \left(\frac{1-a'^2}{a^2}-2\frac{a''}a\right)
    \end{eqnarray}
and
    \begin{eqnarray}
    &&K_{\tau\tau}-Kg_{\tau\tau}=-g^{ij}K_{ij}=
    3\frac{\sqrt{f-a'^2}}a,\\
    &&K_{ij}-Kg_{ij}=\left(2\frac{\sqrt{f-a'^2}}a+\frac1{
    a'}\frac{d}{d\tau}\sqrt{f-a'^2}\right)g_{ij}
    \end{eqnarray}
the Israel junction conditions on the brane become \cite{Kraus}
    \begin{eqnarray}
    &&\frac{\sqrt{f-a'^2}}a=r_c\left(\frac1{a^2}
    -\frac{a'^2}{a^2}\right),                \label{embedding}\\
    &&\frac{d}{d\tau}\Big(a\sqrt{f-a'^2}
    +r_ca'^2\Big)=0,
    \end{eqnarray}
the second of them being just a time derivative of the first one.

Thus, the only independent dynamical equation of brane embedding
reads as (\ref{embedding}) which is exactly the effective Friedmann
equation in the CFT driven cosmology
    \begin{eqnarray}
    \frac1{a^2}
    -\frac{a'^2}{a^2}
    =r_c^2\left(\frac1{a^2}
    -\frac{a'^2}{a^2}\right)^2
    +\frac{\Lambda_5}6+\frac{R_S^2}{a^4}
    \end{eqnarray}
with the identifications of 4D and 5D gravitational and cosmological
constants (\ref{1000})-(\ref{1001}) and the following relation
between the amount of the 4D radiation and the black hole mass
(Schwarzschild radius)
    \begin{eqnarray}
    {\cal C}=R_S^2.                \label{1003}
    \end{eqnarray}

\begin{figure}[h]
\centerline{\epsfxsize 10cm \epsfbox{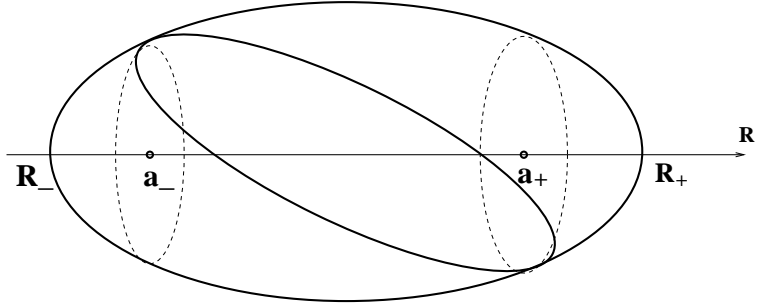}} \caption{\small A
part, $a_-\leq R\leq a_+$, of the 5D Schwarzschild-de Sitter
Euclidean bulk (embedded in the auxiliary 6D spacetime) scanned by
the 4D brane with the induced geometry of the $k=1$ garland
instanton. Every point of this cigar instanton represents {\bf a}
3-sphere of radius $R$. \label{Fig.2}}
\end{figure}

The Euclidean signature of the Schwarzschild-de Sitter metric
(\ref{SdS}) holds for $f(R)>0$ and corresponds to a cigar-like
spacetime of a finite length
    \begin{eqnarray}
    &&R_-<R<R_+,\\
    &&R^2_\pm=\frac1{2H^2}\Big(1\pm\sqrt{1-4H^2{\cal C}}\Big),
    \end{eqnarray}
when considered from the viewpoint of the higher-dimensional
embedding space (see Fig.\ref{Fig.2}). Therefore, in view of
(\ref{apm}) the 4D CFT cosmology instantons with the scale factor
oscillating between $a_\pm$ scan only a part of this cigar 5D
instanton without reaching its poles at $R_\pm$ depicted in
Fig.\ref{Fig.2},
    \begin{eqnarray}
    &&R_-<a_-\leq a<a_+\leq R_+. \label{R_-<R_+}
    \end{eqnarray}

Simultaneously with $R=a(\tau)$ the Euclidean embedding time $T$
evolves in $\tau$ according to (\ref{dTdtau}). This Killing
coordinate of the 5D metric plays the role of the angular coordinate
which should be identified with the period prescribed by the
periodicity of the embedded brane
    \begin{eqnarray}
    \Delta T=\oint d\tau
    \frac{\sqrt{f(a)-a'^2}}{f(a)}=2k\int_{a_-}^{a_+} da
    \frac{\sqrt{f(a)-a'^2}}{\dot a f(a)},
    \end{eqnarray}
where $k$ in the second equality arises for the k-fold garland case
with $k$ oscillations of $a(\tau)$ between its minimal and maximal
values.

\begin{figure}[h]
\centerline{\epsfxsize 10 cm \epsfbox{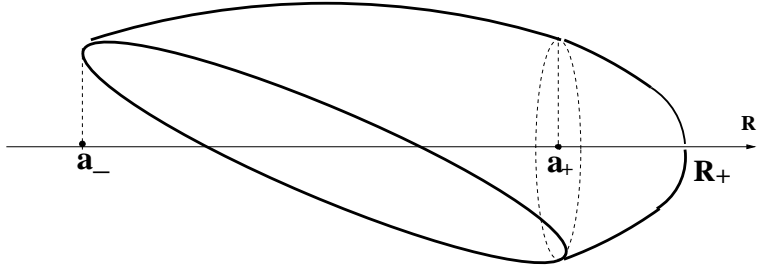}} \caption{\small
Truncation of the 5D Schwarzschild-de Sitter spacetime excluding
from the bulk the Schwarzschild horizon point $R_-$. \label{Fig.3}}
\end{figure}

On the other hand, the period of $T$ dictated by the absence of a
conical singularity at $R_\pm$ equals ($f'(R)\equiv df(R)/dR$)
    \begin{eqnarray}
    \Delta T_\pm=\frac{4\pi}{|f'(R_\pm)|}=
    \frac{\pi\sqrt 2}{H}
    \left(\frac{1\pm\sqrt{1-4R_S^2H^2}}{1-4R_S^2H^2}\right)^{1/2}.
    \end{eqnarray}
As is known \cite{GibHawk,Page} these two periods for the
Schwarzschild-de Sitter solution can be identified only for the
degenerate case of coinciding sizes of Schwarzschild and de Sitter
horizons $R_-=R_+$, which is interpreted as the equilibrium of
Hawking radiation flows emitted by these two horizon surfaces. In
view of (\ref{R_-<R_+}) this is impossible in our case, so that both
conical singularities cannot be simultaneously eliminated. However,
in the presence of the brane (with the $Z_2$-identification across
it) only the region of SdS spacetime $ a(\tau)\leq R\leq R_+$
remains (see Fig.\ref{Fig.3}).

Therefore, a natural selection criterion for the regularity at the
deSitter horizon $R_+$ becomes just one equation relating $\Delta
T_+$ and $\Delta T$
    \begin{eqnarray}
    2k\int_{\tau_-}^{\tau_+} d\tau
    \frac{\sqrt{f(a)-a'^2}}{f(a)}
    =\frac{\pi\sqrt 2}{H}
    \left(\frac{1+\sqrt{1-4R_S^2H^2}}
    {1-4R_S^2H^2}\right)^{1/2}.         \label{cone}
    \end{eqnarray}
This is an equation on ${\cal C}=R_S^2$ in terms of $H^2$ and $B$
analogous to the bootstrap equation (\ref{bootstrap}) on the 4D CFT
side of this problem. It does not however coincide with
(\ref{bootstrap}), so that the suggested DGP/CFT duality does not
incorporate a complete equivalence of 4D and 5D pictures for the
full admissible range of $\Lambda=3H^2$ on the CFT side. If we
insist on the full correspondence of these descriptions, this will
lead to the additional restriction of the $H^2$-range --- the
selection of a finite set of instantons with isolated values of the
cosmological constant. Below we derive those values for static
instantons corresponding to static Einstein universes and for
garlands generated by a conformal scalar field.

\section{Conical singularity in the bulk vs static Einstein universes and
garlands}

For a particular value of the amount of radiation constant --- at
the hyperbolic boundary of the instanton domain in Fig.\ref{Fig.1}
    \begin{eqnarray}
    {\cal C}=\frac{1-2BH^2}{4H^2}         \label{static}
    \end{eqnarray}
the effective equations (\ref{efeq})-(\ref{bootstrap}) have a
one-parameter set of solutions corresponding to static Einstein
universes with a constant $a(\tau)=a_+=a_-=1/H\sqrt2$. Since for
constant $a(\tau)$ the solution can be considered periodic with an
arbitrary period $\eta=\oint d\tau/a=\sqrt2 H\Delta\tau$, the
bootstrap equation (\ref{bootstrap}) does not impose any additional
restriction on the value of $H$ and only determines $H$ in terms of
$\eta$ (or other way around -- $\eta$ in terms of $H$)
    \begin{eqnarray}
    &&2BH^2=\frac1{1+r(\eta)},    \label{Hviar}\\
    &&r(\eta)=\frac2{Bm_P^2}
    \frac{dF(\eta)}{d\eta}=\frac{E_{\rm
    rad}(\eta)}{E_{\rm vac}},              \label{ratio}
    \end{eqnarray}
where $r(\eta)$ is in fact a ratio of the radiation energy to the
vacuum Casimir energy.\footnote{Since $r(\eta)$ is a monotonically
decreasing function of $\eta$, this relation determines a one-to-one
correspondence between $H$ and $\eta$.}

Static solutions exist for any $H$ in the range $0\leq H^2\leq
1/2B$. In particular, for a small $H\to 0$ they have a period
$\eta\to 0$ which in view of the asymptotic behavior
$dF/d\eta\simeq\pi^4/15\eta^4$ (in the case of a scalar field) reads
    \begin{eqnarray}
    \eta\simeq\pi\sqrt{\frac{H}{m_P}}\left(\frac4{15}\right)^{1/4}.
    \end{eqnarray}
This generates the on-shell value of the effective action (which is
given on the static solutions by the Legendre transform of $F(\eta)$
\cite{slih})
    \begin{eqnarray}
    \varGamma_0=F-\eta\frac{dF}{d\eta}\simeq -\frac\pi3
    \left(\frac4{15}\right)^{1/4}\left(\frac{m_P}H\right)^{3/2}
    \to-\infty,\,\,\,\,H\to 0.
    \end{eqnarray}
Therefore these instantons weighted by $\exp(-\varGamma_0)$ are
strongly dominated by the $H\to 0$ limit --- the analogue of the
infrared catastrophe with the Hartle-Hawking no-boundary
wavefunction treated in the tree-level approximation\footnote{Note,
however, that the singularity of $\exp(-\varGamma_0)$ is weaker here
because for the Hartle-Hawking prescription the tree-level action
$-\varGamma_0\sim(m_P/H)^2$ in contrast to $(m_P/H)^{3/2}$ in our
case.}.

This singularity is not dangerous for physical implications in early
Universe, because these static Euclidean solutions are stable and do
not give rise to Lorentzian expanding universes modeling initial
conditions for inflationary cosmology\footnote{For this reason these
solutions were not considered in \cite{slih}.}. They represent
thermodynamic equilibrium ensembles with the temperature
$T=1/\Delta\tau\simeq (15)^{1/4}\sqrt{m_PH}/\pi\to 0$. The same is
true for conformal fields of other spins --- they generate the same
power law behavior for the cosmological parameters, differing only
by spin-dependent coefficients. It is interesting, that the dual 5D
picture of the DGP/CFT correspondence provides a strong truncation
of this family of solutions by selecting just one member from the
range $0\leq H^2\leq 1/2B$.

This truncation follows from requiring the absence of a conical
singularity at the de Sitter horizon of the 5D bulk (\ref{cone}).
Indeed, in the static case (\ref{static}) both sides of this
equation simplify, because $a'=0$, $f(a)=BH^2$ and
$1-4R_S^2H^2=2BH^2$, and it takes the form
    \begin{eqnarray}
    \eta_{\rm cone}=
    \pi\sqrt{2\big(1+\sqrt{2BH^2}\big)}.      \label{conelesseta}
    \end{eqnarray}
The substitution of this expression to (\ref{Hviar}) gives a closed
equation for {\bf the} numerical value of $2BH^2$. This equation always has
a solution between zero and one. Note that the period for the
coneless case has a very limited range $\pi\sqrt2\leq\eta\leq 2\pi$
in which the evaluation of $F(\eta)$, (\ref{freeenergy}), with a big
accuracy is given by the contribution of several lowest oscillator
levels $\omega$. The numerical solution of this equation does not
present any difficulty, and it gives, for a pure scalar field with
$Bm_P^2=1/240$
    \begin{eqnarray}
    2BH^2\simeq 0.3612,\,\,\,
    r\simeq 1.7688,                \label{coneless0}
    \end{eqnarray}
(the corresponding thermal sum of (\ref{bootstrap}) runs over the
eigenfrequencies on a unit 3-sphere $\omega_n=n$ with weights
$\sum_\omega\equiv\sum_{n\geq 1} n^2$).

For a pure 2-component Weyl spinor field with $Bm_P^2=11/480$,
$\omega_n=n+1/2$ and $\sum_\omega\equiv\sum_{n\geq 1} n(n+1)$ we
have
    \begin{eqnarray}
    2BH^2\simeq 0.9858,\,\,\,
    r\simeq 0.0144,                \label{coneless1/2}
    \end{eqnarray}
and for a pure vector field with $Bm_P^2=31/120$,
$\omega_n^2=2(n^2-2)$ and $\sum_\omega\equiv 2\sum_{n\geq 2}(n^2-1)$
    \begin{eqnarray}
    2BH^2\simeq 0.9998,\,\,\,
    r\simeq 0.00016\ll 1.               \label{coneless1}
    \end{eqnarray}
Interestingly, for spinors and vectors the cone free instantons are
very close to the upper bound of the range (\ref{landscape}) --- the
new quantum gravity scale $H^2=1/2B$.

When the CFT is composed of many fields of one spin, these values do
not depend on their number $N$ (because $N$ cancels out in the ratio
(\ref{ratio}). In the case of the mixture of spins, they depend on
the partial weights of these spins in the full system. Then the
resulting value of $2BH^2$ lies somewhere between (\ref{coneless0})
and (\ref{coneless1}).

Thus, only one member of the static Einstein family with
$\Lambda\neq 0$ remains in a dual DGP/CFT description, which again
eliminates the infrared catastrophe of $\Lambda\to 0$ (even though
this catastrophe for this family is irrelevant from the viewpoint of
early cosmology). Application of the same criterion of dual
description to garland families by the combination of
Eqs.(\ref{cone}) and (\ref{bootstrap}) also restricts them to a
finite set (intersection of the one-parameter set of solutions of
(\ref{cone}) with one-parameter sets of garlands). In view of the
irreducible integral nature of these equations this can hardly be
done analytically. Numerical calculations show that for the scalar
field there is a single member of the $k=2$ garland family with the
following value of the Hubble constant and the amount of radiation
constant
    \begin{eqnarray}
    H^2_{\rm garland}\simeq 56.5\,m_P^2,\,\,\,
    {\cal C}=0.0018\,m_P^{-2}.                \label{conelessgarland}
    \end{eqnarray}
In fact this instanton is rather close to the static instanton of
(\ref{coneless0}).

For other spins and their mixtures we do not have concrete results,
but it is expected that a dual description also exists for a finite
set of garlands with isolated values of the cosmological constant.
This  problem will be analyzed elsewhere. In particular, it is
anticipated that, when vector and spinor contributions dominate in
the particle phenomenology, the DGP/CFT duality will enforce the
values of the cosmological and amount of radiation constants close
to (\ref{coneless1/2})-(\ref{coneless1}), that is very close to the
upper boundary of the range (\ref{landscape}).

This duality is very much equivalent to the AdS/CFT correspondence
picture not only from the viewpoint of classical-to-quantum
correspondence, but also from the viewpoint of energy scales of the
problems. Indeed, on the CFT cosmology side the limit $2BH^2\simeq
1$ corresponds to the uppermost possible value of $\Lambda$. With
the number of conformal fields $N\gg 1$ the corresponding
$\Lambda/3\simeq 1/2B\sim m_P^2/N$ is small, but it still turns out
to be in a strong coupling limit --- at the cutoff generated in
gravitational physics by a large number of quantum species
\cite{species1,species2}. On the 5D side, however, we have for large
$N$ large semiclassical black holes. This follows from the fact that
for large $B\sim N$ the black hole horizon in the vicinity of the
family of solutions (\ref{static}) can be estimated as
    \begin{eqnarray}
    R^2_S={\cal C}\simeq
    \frac{B}2\, r\sim \frac{N}{m_P^2}\,r\gg m_P^{-2},   \label{classicalBH}
    \end{eqnarray}
because the radiation to vacuum energy ratio (\ref{ratio}) is a
quantity essentially independent of the total $N$. Thus, this is a
nearly classical black hole provided the number of species is larger
than $1/r$. Moreover the dynamics of the brane cosmology takes place
at $a_\pm^2=1/2H^2\simeq B\sim N/m_P^2$ far away from the BH
horizon, where quantum effects of the bulk background are even
weaker.

\section{Towards background independent duality}
The duality between the 4D CFT setting and 5D braneworld setup holds
in a cosmological sector, with degrees of freedom which are
spatially homogeneous collective variables. The duality between both
models is far from being obvious when those degrees of freedom are
not restricted to be homogeneous, as is the case for the background
cosmology, but are also free to contain spatially inhomogeneous
modes. This can be checked only by comparing relevant sets of
correlation functions considered within perturbation theory on this
background. However, the duality construction of the above type can
be considered as a step towards {\em background independent}
classical-to-quantum correspondence between the theories in
neighboring spacetime dimensions \cite{DvaliGomez}. Our case goes
beyond a conventional AdS/CFT correspondence, where the bulk
Schwarzschild-AdS background and its asymptotically remote boundary
are in fact non-dynamical (being a trivial static solution of 10D
SUGRA equations parameterized by fixed coupling constants). Here the
4D thermal brane is located at a finite variable distance from the
BH and undergoes a nontrivial evolution in the bulk background. The
latter is static but involves the BH mass (and the relevant amount
of radiation on the brane) which is not a free input but rather a
quantity fixed by nontrivial consistency conditions.

Moreover, this model contains the indication that this duality can
be extended further, beyond any assumptions on spacetime symmetries
of the underlying background. As we will see, a part of the full
nonlinear equations in both settings of this model coincide
irrespective of the particular symmetry ansatz for the physical
configurations. To show this, consider the Einstein equations with
Israel boundary (junction) conditions
(\ref{bulkEinstein})-(\ref{surfacestress}). With the DGP scale
parameter (\ref{DGPscale}) the junction conditions can be rewritten
as
    \begin{eqnarray}
    K_{\mu\nu}=-r_c\Big(
    R_{\mu\nu}
    -\frac16 \,g_{\mu\nu}\,R \Big),
    \end{eqnarray}
and the gravitational constraint in the bulk -- the normal-normal
projection of (\ref{bulkEinstein}),
    \begin{eqnarray}
    &&-2\,n^A n^B\Big(R^{(5)}_{AB}-\frac12\,G_{AB}\,R^{(5)}
    +\Lambda_5\,G_{AB}\Big)\nonumber\\
    &&\qquad\qquad\qquad\qquad\qquad\quad
    =R^{(4)}+K_{\mu\nu}^2-K^2-2\Lambda_5=0,
    \end{eqnarray}
takes the form of the following equation entirely in terms of 4D
brane quantities
    \begin{eqnarray}
    R^{(4)}+r_c^2
    \Big(R_{\mu\nu}^2-\frac13\,R^2\Big)-2\Lambda_5=0, \label{5Dside}
    \end{eqnarray}
because the quadratic form in the extrinsic curvature reads as
$K_{\mu\nu}^2-K^2=r_c^2(R_{\mu\nu}^2-R^2/3)$ -- a structure
related by a total derivative term to the square of the Weyl tensor.
Thus, unexpected conformal properties start popping out in this
seemingly unrelated to conformal invariance braneworld setup. The
{\em tree-level} equation (\ref{5Dside}) is exact and totally
independent of any background ansatz restrictions.

Let us now consider the {\em quantum effective} equations on the 4D
CFT side. Without any symmetry assumptions and selection of
collective variables the full effective action in the theory (in the
$1/N$-approximation) can be represented as a sum of Einstein term,
the conformal anomaly action $\varGamma_A$ and conformally invariant
part including the action for thermal radiation and vacuum (Casimir)
energy $\varGamma_{T,C}$
    \begin{eqnarray}
    \varGamma=
    -\frac1{16\pi G_4}\int d^4x\,
    g^{1/2}\,\Big(R^{(4)}-2\Lambda_4\Big)
    +\varGamma_A+\varGamma_{T,C}.
    \end{eqnarray}

Now consider the conformal variation of this action or the trace
part of effective Einstein equations. We will assume that, as
before, the number of degrees of freedom preserving renormalization
has been done, so that the conformal variation of the anomalous
action has the form (\ref{anomaly}) with the coefficient of the
$\Box R$ term $\alpha=0$. With a known equation for the Gauss-Bonnet
invariant $E = R_{\mu\nu\alpha\gamma}^2-4R_{\mu\nu}^2 + R^2$ in
terms of the Weyl tensor
    \begin{eqnarray}
    E =C^2_{\mu\nu\alpha\beta}
    -2\Big(R_{\mu\nu}^2-\frac13\,R^2\Big)        \label{Weyl-Euler}
    \end{eqnarray}
this anomalous variation reads
    \begin{eqnarray}
    g_{\mu\nu}\frac{\delta
    \varGamma_A}{\delta g_{\mu\nu}} =
    \frac{1}{64\pi^2}g^{1/2}
    \left[-2\beta \Big(R_{\mu\nu}^2-\frac13\,R^2\Big)+
    (\beta+\gamma) C_{\mu\nu\alpha\beta}^2\right].    \label{anomaly1}
    \end{eqnarray}
Therefore, in view of the local conformal invariance of
$\varGamma_{T,C}$ the full trace effective equation
$g_{\mu\nu}\delta \varGamma/\delta g_{\mu\nu}=0$ takes the form
    \begin{eqnarray}
    R+\frac{\beta
    G_4}{2\pi}\Big(R_{\mu\nu}^2
    -\frac13\,R^2\Big)-4\Lambda_4
    -\frac{\beta
    +\gamma}{4\pi}\,G_4\,C^2_{\mu\nu\alpha\beta}=0.   \label{4Dside}
    \end{eqnarray}

Under the known indentifications of the 4D and 5D parameters
(\ref{1000})-(\ref{1001}), the first three terms here exactly
reproduce the equation (\ref{5Dside}) on a 5D braneworld side -- the
tree-level generalized DGP model. For a cosmological setting with a
conformally flat FRW metric the last term is vanishing, but for a
generic field configurations it is nonzero, and these 5D and 4D
pictures are generally not dual. However, salvation might come from
the supersymmetry requirement. It is a well-known fact that in the
superconformal field theory --- the $D=4$ $SU(\mathbb{N})$ SYM with
an extended ${\cal N}=4$ supersymmetry
\cite{Duffanomaly,TseytlinLiu}, which is a typical object of AdS/CFT
correspondence with the particle contents
$(N_0,N_{1/2},N_1)=(6\mathbb{N}^2,4\mathbb{N}^2,\mathbb{N}^2)$, cf.
Eq.(\ref{100}),
--- the coefficients of the conformal anomaly satisfy the relation
    \begin{eqnarray}
    \gamma+\beta=0.
    \end{eqnarray}
This renders equations (\ref{5Dside}) and (\ref{4Dside}) completely
equivalent for a generic metric configuration. In particular, the
particle model with the same phenomenology\footnote{One should bear
in mind that this will not be literally the SCFT of the above type,
because of a positive cosmological constant.} when applied in our
CFT cosmology will provide a hierarchy of classical-to-quantum
coupling scale limits (\ref{classicalBH}) with $N\sim\mathbb{N}^2\gg
1$.

One can find for this particle content the static instanton (on the
hyperbolic curve of Fig.\ref{Fig.1}) satisfying the selection
criteria on both 4D and 5D sides. According to the previous section,
it represents a numerical solution of Eq.(\ref{Hviar}) with the
conformal time (\ref{conelesseta}) for the above set of spins with
the relevant value of the total $Bm_P^2
=(6B_0+4B_{1/2}+B_1)m_P^2\mathbb{N}^2 =3\mathbb{N}^2/8$ and the
relevant thermal scalar, spinor and vector sums  weighted in
$dF(\eta)/d\eta$ by the numbers of particles $N_s$, cf.
Eq.(\ref{bootstrap}). It turns out that this static instanton has
the parameters
    \begin{eqnarray}
    2BH^2\simeq 0.937,\,\,\,
    r\simeq 0.0673,               \label{conelessSCFT}
    \end{eqnarray}
which again nearly saturate the cutoff bound $H^2=1/2B$. Apparently,
the garland instanton satisfying the same criteria of a complete
duality is also very close to this point, which supports the
conclusions of Sect.5 on the weak coupling to strong coupling
correspondence of relevant 5D and 4D pictures.

\section{The case of a negative cosmological constant}

The previous section gives a strong motivation for the introduction
of supersymmetry in the particle phenomenology of the model, which
however requires negative values of the bulk and brane cosmological
constants in (\ref{1001}). For this reason let us consider the case
of
    \begin{eqnarray}
    \Lambda_4\equiv-3H^2<0.   \label{negative}
    \end{eqnarray}
We will see that this results in the flip of the sign of the {\em
effective} cosmological constant, which makes this choice useful
also for the use in inflation theory.

With this choice the formal turning points (\ref{apm}) go over into
    \begin{eqnarray}
    a_\pm^2=\frac{\pm\sqrt{1+2BH^2+4{\cal C}H^2}-1}{2H^2},
    \end{eqnarray}
and in view of $a_-^2<0$ the Euclidean evolution of the scale factor
can only be in the range $0\leq a(\tau)\leq a_+$, which corresponds
to the formation of the Hartle-Hawking vacuum instanton with the
$S^4$ topology -- degenerate case of a torus ripped at the vanishing
value of the scale factor. The Euclidean evolution domain
$a(\tau)\geq a_+$ is of no interest, because it is not periodic and
of no relevance to the statistical sum of the model.

With $a=0$ at the boundary of the evolution domain the conformal
time diverges to infinity, $\eta\to\infty$, so that
$dF(\eta)/d\eta\to 0$, and ${\cal C}=0$ from (\ref{bootstrap}). Then
the effective Friedmann equation (\ref{efeq}) reduces to
    \begin{eqnarray}
    &&a'^2=1-a^2\frac{1\pm\sqrt{1+2BH^2}}B
    =1-H^2_{\rm eff}a^2,                        \label{50}\\
    &&H^2_{\rm eff}=\frac{1+\sqrt{1+2BH^2}}B,
    \end{eqnarray}
where the plus sign of the square root in the expression for $a'^2$
is chosen to have a necessary turning point at $a_+=1/H_{\rm eff}$
--- the equatorial section of the spherical Hartle-Hawking instanton
mentioned above. The solution of (\ref{50}) is an obvious Euclidean
de Sitter metric with {\bf a} {\em positive} effective cosmological
constant $\Lambda_{\rm eff}=3H^2_{\rm eff}$ (cf. Eq.(\ref{HHHeff})
for the Hartle-Hawking instanton in the case of a positive
primordial $\Lambda=3H^2$). When rewritten in terms of the original
negative cosmological constant (\ref{negative}) this effective one
explicitly features the sign flip
    \begin{eqnarray}
    \Lambda_{\rm eff}
    =-\frac{2\Lambda_4}{\sqrt{1-2B\Lambda_4/3}-1}>0.
    \end{eqnarray}
This nonlinear transition from the negative primordial cosmological
constant to the positive effective one is a genuine quantum effect
of the conformal anomaly. This effect, for example, makes
unnecessary the KKLT uplifting \cite{KKLT} of the negative vacuum
energy --- here the primordial cosmological constant can be negative
and consistent with supersymmetry and still can generate a positive
effective $\Lambda_{\rm eff}$ needed for inflation.

Unfortunately, there is an obstacle to the realization of this
elegant mechanism. If we believe in the microcanonical initial
conditions of the above type, then the contribution of the
Hartle-Hawking instanton should be weighted by the exponentiated
negative of the Euclidean effective action. But, as mentioned above,
the latter diverges to $+\infty$ \cite{slih,why} by virtue of the
conformal anomaly part and suppresses to zero this contribution of a
primordial $\Lambda_4<0$. There is a hope to overcome this
difficulty by using the tunneling prescription
\cite{tunnel,Vilenkin} alternative to the no-boundary construction
of the above type. As suggested in \cite{TQC} the tunneling state
can be incorporated into the microcanonical path integral
formulation in the form of another saddle point of (\ref{1}) with
the {\em negative} value of the lapse function $N=-1$. In the main
it leads to the reversal of the sign of the semiclassical
exponential in the distribution function of the quantum ensemble.
But such a tunneling state was calculated in \cite{TQC} only for a
system with non-conformal quantum fields with heavy masses,
admitting a local gradient expansion of the effective action. The
inclusion of massless conformal fields into this construction
encounters certain difficulties and is currently under study.

\section{Conclusions}

Thus we have constructed a 5D braneworld setup dual to the
microcanonical state of a 4D spatially closed cosmology driven by
the conformal field theory with a large number of quantum fields.
The importance of this model lies in the fact that it provides
initial conditions for inflationary cosmology in a limited range of
the effective cosmological Hubble factor, which at late stages of
expansion can also have a stage of the cosmological acceleration by
the big boost scenario of Ref. \cite{boost}. The dual 5D picture is
represented by the generalized DGP model with no matter on the
brane, but with a positive primordial cosmological constant in the
bulk and a black hole imitating the radiation of conformal fields on
the cosmological 4D CFT side. For a large number of quantum CFT
species $N$ the DGP/CFT duality holds for distinguished values of
the cosmological constant, which correspond to the classical weak
coupling regime of the 5D Schwarzschild-de Sitter bulk and a strong
coupling regime of the 4D CFT at the natural cutoff induced by large
$N$.

This duality is fully proven for global cosmological degrees of
freedom of the system, and its extension to the sector of spatially
inhomogeneous modes and their perturbative correlators still has to
be done. However, there is an indication that this can be done even
at the level of the full nonlinear equations, because a trace part
of effective Einstein equations exactly coincides on both sides of
the duality relation for a generic metric configuration unrestricted
by any Killing symmetry considerations. This only imposes
constraints on particle phenomenology provided, in particular, by
superconformal symmetry of the theory. This suggests a new type of
background independent classical-to-quantum correspondence. Finally,
we proposed a  mechanism for flipping the sign of the effective
cosmological constant in the CFT driven cosmology, which can
reconcile the negative, compatible with supersymmetry, value of the
primordial cosmological constant with inflationary cosmology.

\section*{Acknowledgements}
A.B. is grateful for hospitality of the Laboratory APC in Paris
(CNRS, Paris VII University, CEA, Paris Observatory) and the
Laboratory MPT CNRS-UMR 6083 of the University of Tours, where a
major part of this work has been done. His work was also supported
by the Russian Foundation for Basic Research under the grant No
08-01-00737 and the grant LSS-1615.2008.2. A.K. was supported by the
RFBR grant 08-02-00725 and the grant LSS-4899.2008.2. A.B. wishes to
thank G.Dvali, P.Mazur, E.Mottola, M.Shaposhnikov, S.Solodukhin,
A.Starobinsky and R.Woodard for discussions.

\end{document}